\documentclass[epj,twocolumn]{webofc}
\usepackage[varg]{txfonts}   % Web of Conferences font
\usepackage{graphicx}

\wocname{epj}
\woctitle{Seismology of the Sun and the Distant Stars 2016}
\begin{document}
%***********************************************************************
\title{The additional-mode garden of RR Lyrae stars}

\author{\firstname{L\'aszl\'o} \lastname{Moln\'ar}\inst{1}\fnsep\thanks{\email{molnar.laszlo@csfk.mta.hu}} \and
        \firstname{Emese} \lastname{Plachy}\inst{1} \and
        \firstname{P\'eter} \lastname{Klagyivik}\inst{1} \and
        \firstname{\'Aron L.} \lastname{Juh\'asz}\inst{1,2} \and
        \firstname{R\'obert} \lastname{Szab\'o}\inst{1} \and
        \firstname{Zachary} \lastname{D'Alessandro}\inst{3} \and
        \firstname{Benjamin} \lastname{Kratz}\inst{3} \and
        \firstname{Justin} \lastname{Ortega}\inst{3} \and
        \firstname{Shashi} \lastname{Kanbur}\inst{3}
        % etc.
}

\institute{Konkoly Observatory, MTA CSFK, Konkoly Thege Mikl\'os \'ut 15-17, H-1121 Budapest, Hungary
\and
           E\"otv\"os Lor\'and University, P\'azm\'any P\'eter s\'et\'any 1/a, H-1117 Budapest, Hungary 
\and
           State University New York, 7060 Route 104, Oswego, NY 13126-3599, USA
          }

%-----------------------------------------------------------------------
\abstract{%
Space-based photometric missions revealed a surprising abundance of millimagnitude-level additional modes in RR Lyrae stars. The modes that appear in the modulated fundamental-mode (RRab) stars can be ordered into four major categories. Here we present the distribution of these groups in the Petersen diagram, and discuss their characteristics and connections to additional modes observed in other RR Lyrae stars.
}
\maketitle
%
%-----------------------------------------------------------------------
\section{Introduction}
\label{intro}
RR Lyrae stars have been classified into three main categories based on the pulsation modes present in their light variations. These classical Bailey classes are designated as RRab, RRc, and RRd for the fundamental-mode, first-overtone and double-mode stars, respectively. (A fourth, second-overtone group is claimed to exist in the Magellanic Clouds, but these RRe stars are yet to be found in the Milky Way \cite{2009AcA....59....1S}.) However, the discoveries of the past decade, mainly from the OGLE survey and the photometric space missions, complicated this simple picture. The RRc population, for example, has been extensively studied, and a new family of modes called 0.61-type, or $f_X$ modes was discovered. These modes are strongly connected to the first radial overtone, and can be detected in both first-overtone and double-mode RR Lyrae and Cepheid stars. These new results strongly suggest that RRc and RRd stars should be considered as stars dominated by one or two radial modes, but are not, in fact, single- and double-mode stars.

Also, it seems that at least four main groups can be recognized, by splitting the RRab group into two subgroups: the truly single-mode, and the modulated-multimode types \cite{2016CoKon.105...11M}.  According to the results of \textit{Kepler} and \textit{CoRoT}, non-Blazhko RRab stars seem to be pure radial pulsators with no indications of other modes down to sub-mmag levels. In contrast, modulated RRab stars exhibit a variety of low-amplitude additional modes (see, e.g., \cite{2015ApJ...809L..19B}). However, the distribution of these modes was not investigated yet in detail.

In this paper we present our first results from a systematic survey of additional modes in RRab stars, based mostly on the early campaigns of the K2 mission of the \textit{Kepler} space telescope. Here we only discuss the new landscape and identify the main groups of the new modes. Detailed findings, including the amplitudes and tabulated values will be presented in a forthcoming paper. Another restriction is that here we only focus on modes that appear at frequencies between $f_0$ and $2f_0$, e.g., we do not include ones that could be \textit{g}-modes at low frequencies or higher-order $p$-mode overtones \cite{2015MNRAS.452.4283M,2012MNRAS.427.1517S}. The only exception is period doubling:  the strong 9:2 resonance that destabilises the limit cycle of the fundamental mode and causes it to bifurcate, occurs between the fundamental mode and the 9th radial overtone \cite{2011MNRAS.414.1111K}. However, because of the resonance, the mode itself is usually not directly observable, but the strongest signature of the bifurcation, the half-integer peak at $\approx 3/2 f_0$, appears within the search range of our investigation. 

%++++++++++++++
\begin{figure}
\centering
\includegraphics[width=\hsize]{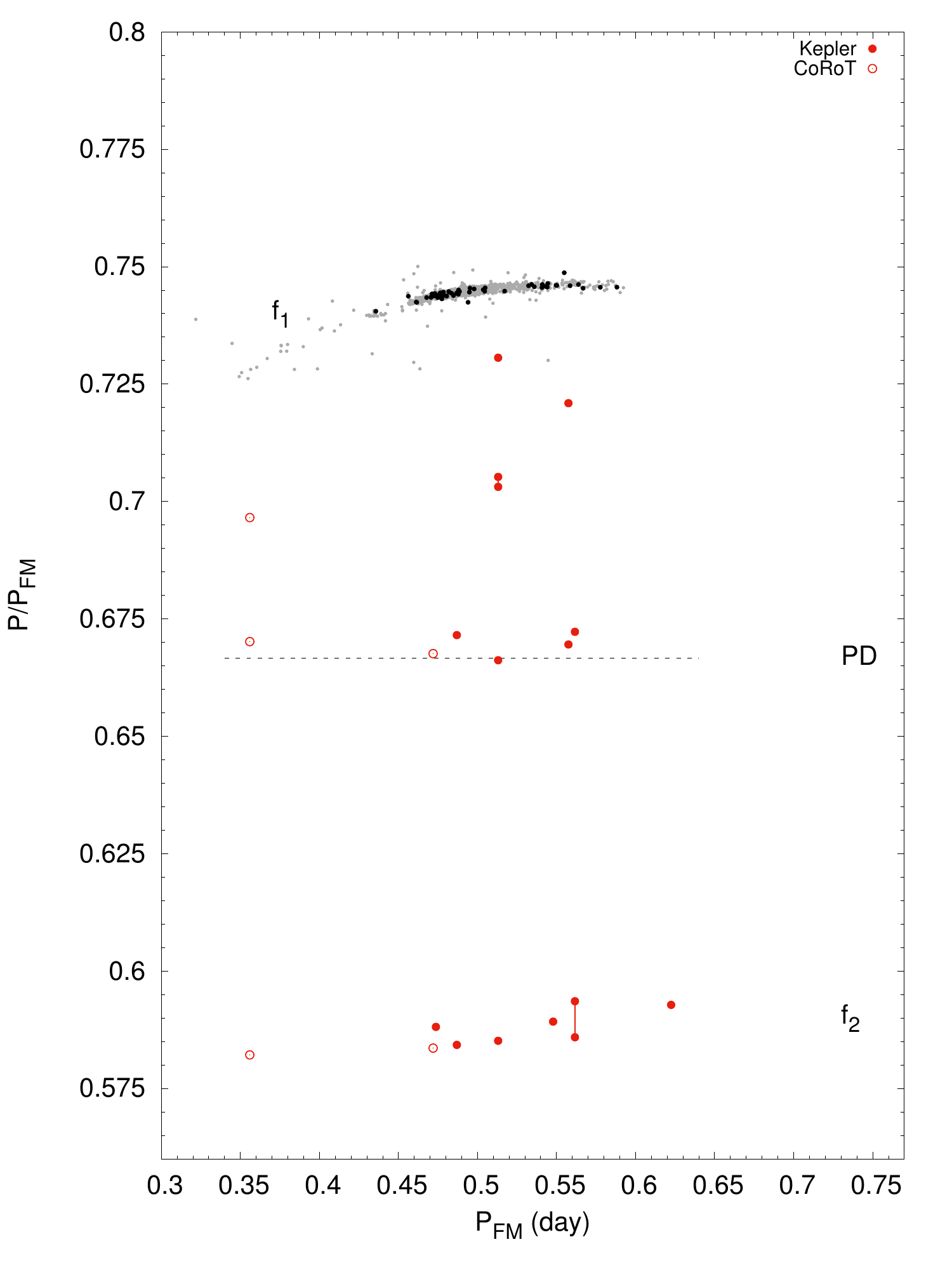}
\caption{Petersen diagram of the first additional modes detected in fundamental-mode stars, compared to the distribution of classical double-mode stars (grey and black). The dashed line marks the 2/3 ratio where the half-integer peaks of period doubling are expected.}
\label{fig-1}       % Give a unique label
\end{figure}
%++++++++++++++

%-----------------------------------------------------------------------
\section{Additional modes in RRab stars}
\label{sec-1}

The first additional modes in RRab stars were discovered by the observations of the \textit{CoRoT} space telescope \cite{2010A&A...510A..39C,2011MNRAS.415.1577G}. In parallel, the first observations of the \textit{Kepler} space telescope also revealed peaks in multiple stars at a a level of 1-10 mmag in the Fourier spectra \cite{2010MNRAS.409.1585B}. These first detections are summarised in the Petersen diagram in figure~\ref{fig-1}. The 
small grey and black dots represent the classical RRd stars in the OGLE-III sample (Magellanic Clouds and the bulge) and the Galactic field, respectively \cite{2009AcA....59....1S,2011AcA....61....1S,2010AcA....60..165S}. Two groups can be readily identified in the Petersen diagram. One is at P/P$_0$ = 2/3, or at the inverse of the half-integer frequency peak: these are the period-doubled (PD) stars. The other group appears around P/P$_0 \approx 0.59$. This value agrees with the expected range of the second radial overtone (P$_2$/P$_0$ = 0.55\dots0.62), hence it is commonly labeled as the $f_2$ group, even though the current one-dimensional, non-linear hydrodynamical models cannot generate second-overtone pulsations to confirm the nature of this group. The rest of the modes spread between the half-integer ones and the position of the first overtone, labeled as $f_1$. 

%++++++++++++++
\begin{figure}
\centering
\includegraphics[width=\hsize]{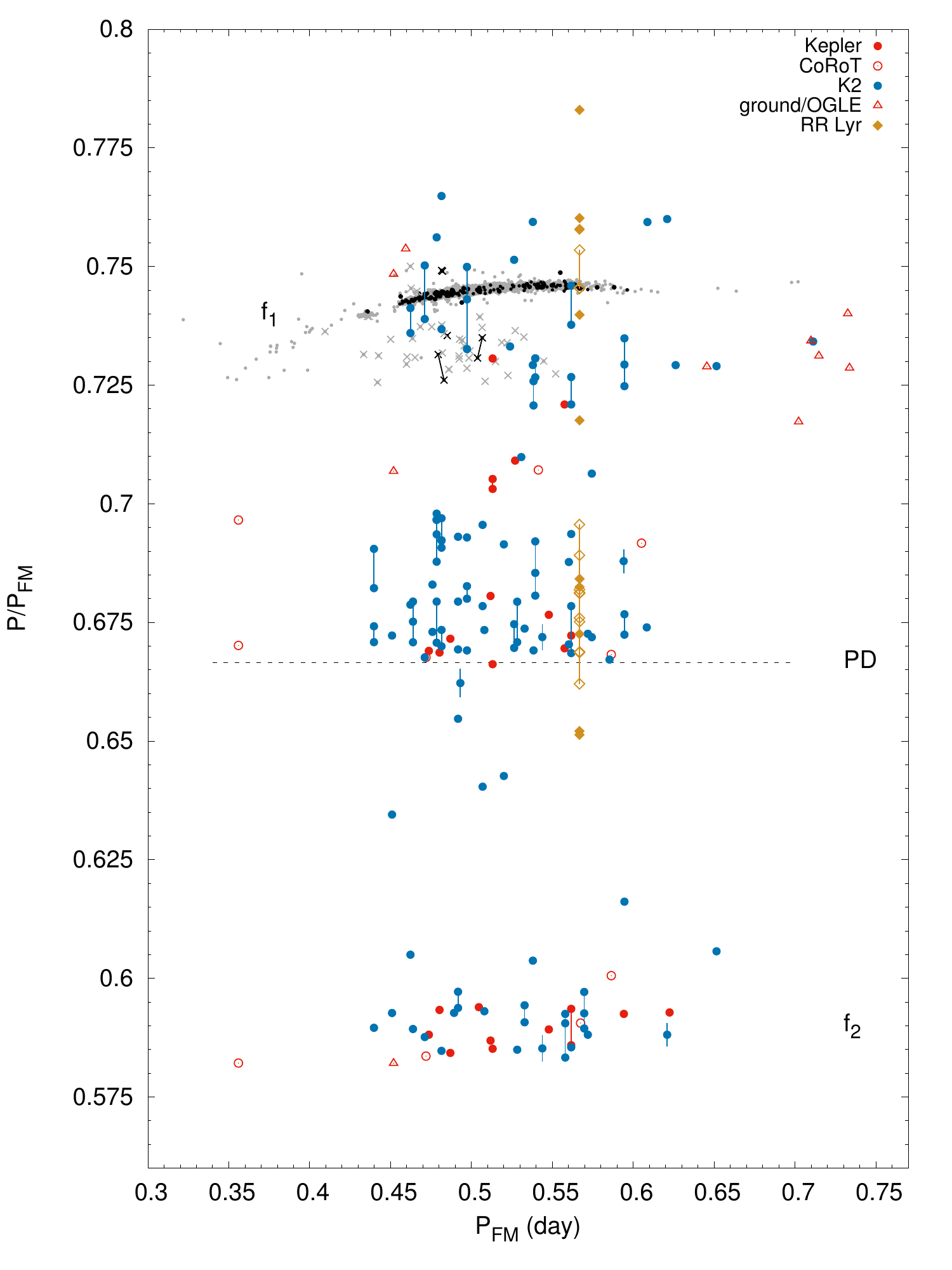}
\caption{Petersen diagram of RRab stars, with the K2-E, K2 campaigns 0--3 and 6, and other data sets included. See the text for more details.}
\label{fig-2}       % Give a unique label
\end{figure}
%++++++++++++++

%.......................................................................
\subsection{The current sample of additional modes}
\label{sec-2}
During the six years that passed since the first additional modes described in V1127 Aql, the sample has expanded greatly, mostly due to the efforts of the K2 RR Lyrae Survey (Szab\'o et al., in these proceedings). One important result came from the star RR Lyrae itself. The continuous \textit{Kepler} light curve of the star revealed that not only a strongly variable period doubling, but another mode is also present in the star that could be the first radial overtone \cite{2012ApJ...757L..13M}. A reanalysis of the \textit{Kepler} quarters 5, 6, 13, and 14 uncovered a complicated structure of peaks at the regions of both the half-integer peak and the first overtone. These are marked with golden diamonds in figure~\ref{fig-2}. Here the empty diamonds mark peaks that are separated approximately by the modulation frequency of RR Lyr, suggesting that these modes also experience the Blazhko effect to at least some extent. 

Additional modes were identified in a few more stars that were observed by \textit{CoRoT}, the Konkoly Blazhko Survey, OGLE-IV and from Antarctica \cite{2012MNRAS.427.1517S,2014A&A...570A.100S,2016MNRAS.461.2934S,2016MNRAS.456..192Z,2016MNRAS.tmp.1249B}, but most of the new detections originate from the K2 mission. This is due to the large field-of-view and aperture of \textit{Kepler}, the step-and-stare approach of the mission, and the plethora of newly discovered RR Lyrae stars that can be targeted, discovered mainly by the LINEAR and Catalina Sky Surveys \cite{2014ApJS..213....9D,2013AJ....146...21S}. For this initial study, we analysed the frequency contents of 307 RRab stars observed in K2 Campaigns 0-3 and 6, and included the results from the 27 stars detected in the K2-E2 engineering run \cite{2015MNRAS.452.4283M}. The K2 observations were reduced with the Extended Aperture Photometry method (Plachy et al., these proceedings), and the pulsation frequencies were determined with the Period04 software \cite{2005CoAst.146...53L}. Overall, we detected additional modes in 40 stars, e.g., 12\% of the K2 sample.

\begin{figure}
\centering
\includegraphics[width=\hsize]{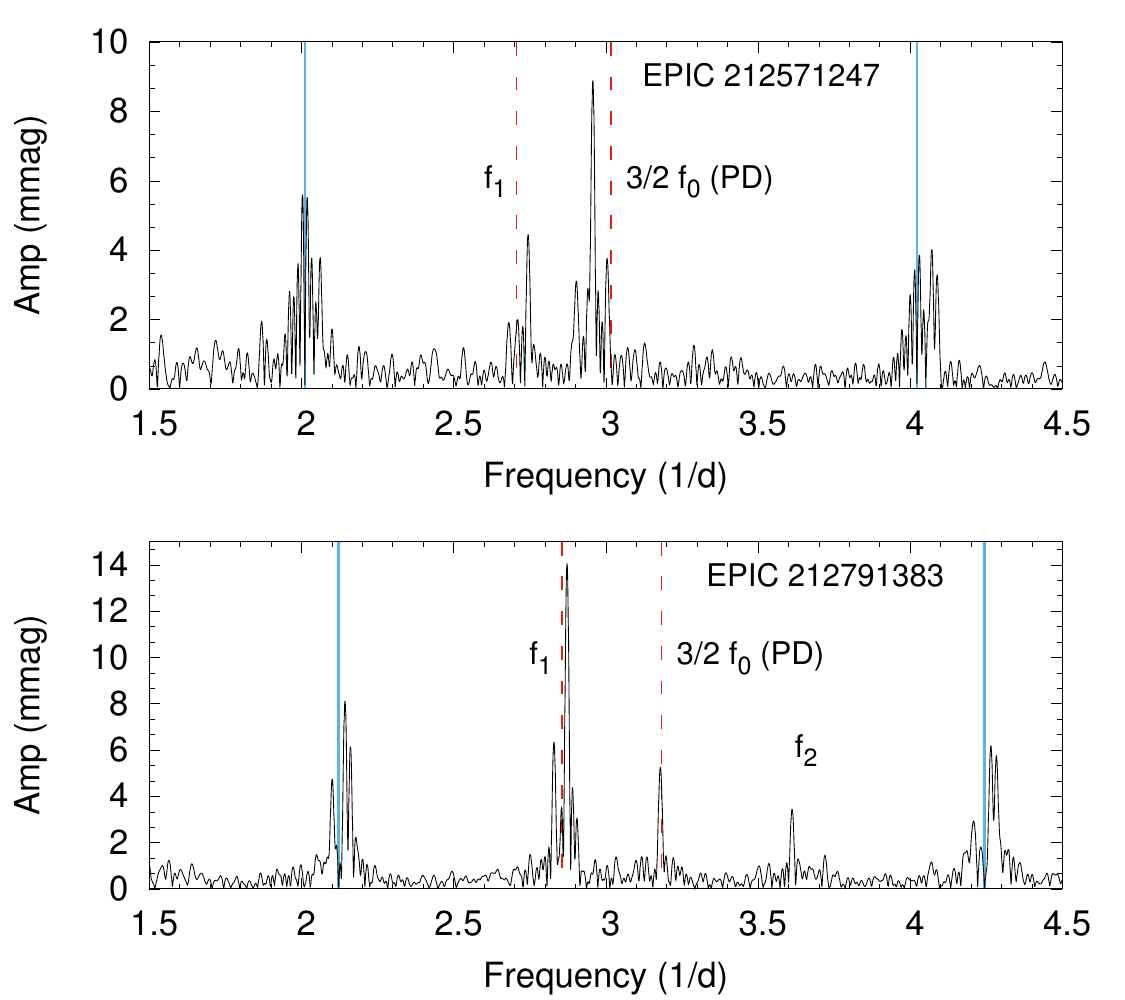}
\caption{Two examples from Campaign 6. Upper panel: note the offset from canonical values (red dashed lines). Lower panel: a rare example with all three types of additional modes present. Pulsation peaks (blue lines) have been prewhitened.}
\label{fig-3}       % Give a unique label
\end{figure}

The investigation of the K2 data revealed interesting patterns in the occurrence of these additional modes. In most cases, only modes belonging to one or two main groups could be identified. A rare example with all three types ($f_1$, $f_2$, PD) present in the star simultaneously, is shown in the lower panel of figure~\ref{fig-3}. Another feature we observed in many stars is when multiple peaks are present, the strongest ones in the groups are not necessary the central peaks, and/or the strongest peaks are offset from the expected, canonical values of period doubling and/or the first overtone. The upper panel of figure~\ref{fig-3} clearly shows these signatures. The $f_1$ group is centered at the canonical value, but the highest-amplitude one is the one of highest frequency; whereas the PD group is shifted from the 3/2 $f_0$ value, but it is symmetric in terms of amplitudes. These details about the RRab additional modes are currently not understood.

%++++++++++++++
%++++++++++++++

For completeness, we also included the new, anomalous RRd stars in figure~\ref{fig-2} with small crosses: grey crosses are OGLE-IV stars that feature either peculiar period ratios or modulations or both \cite{2016AcA....66..131S}, whereas black crosses are the Blazhko-RRd stars discovered in the globular cluster M3 \cite{2014ApJ...797L...3J}.

\subsection{Distribution of the modes by period ratios}
The main aim of the study was to identify sub-structures in the distribution of the additional modes, just like the three distinct groups found in the Petersen diagram of RRc stars \cite{2015MNRAS.453.2022N}. As figure~\ref{fig-2} illustrates, this was not fully realised. The $f_2$ group remained a broad but distinct feature at the bottom of the Petersen diagram. Upwards, a few points show up in the previously empty range between P/P$_0$ = 0.60\dots0.66. 

The most dramatic change can be seen at and above the expected position of the peaks of period doubling. Here, in the middle of the figure, we no longer see a definite ridge along the P/P$_0$ = 2/3 line. The points spread upwards, to P/P$_0$ = 0.70. We suspect that in several cases, the temporal variations of either the additional mode or the fundamental mode, or both, give rise to multiple peaks. Some obvious cases are linked with lines in the figure. However, this does not explain the occurrence of all peaks. It seems that bona-fide period doubling (resonant interaction with the 9th overtone) may suffer confusion with other modes that are excited in the P/P$_0$ = 0.66\dots0.70 range. 

Finally, by now, multiple stars can be identified in which the additional modes overlap with region of the first overtone in RRd stars, especially with the ones with anomalous period ratios. The existence of this group can be described by pure radial pulsations: hydrodynamic models showed that once the period-doubling bifurcation occurs, the new limit cycle can become unstable against the first overtone. These triple-mode models are unlike RRd stars: they are still dominated by the fundamental mode, and may occur over a broader parameter regime, and the interaction between the modes can lead to the appearance of resonant and chaotic states \cite{2013MNRAS.433.3590P}. 

%++++++++++++++
\begin{figure}
\centering
\includegraphics[width=\hsize]{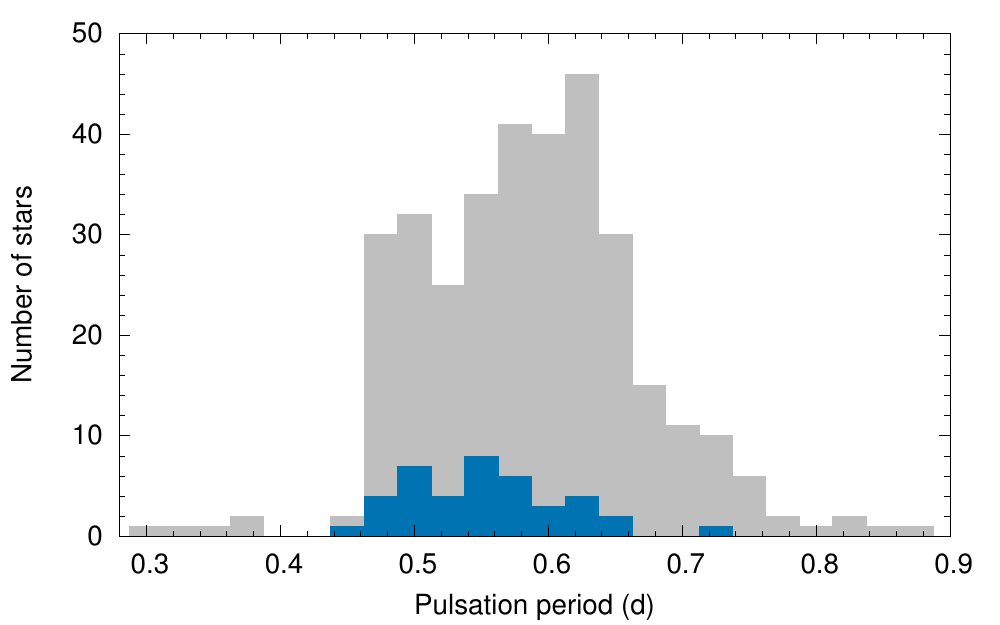}
\caption{In grey: the period distribution of all known RRab stars in the K2 sample investigated here: In blue: the distribution of stars where addtional modes were identified. }
\label{fig-4}       % Give a unique label
\end{figure}
%++++++++++++++

\subsection{Distribution by pulsation periods}
Interestingly, all but one RRab stars with additional modes from the K2 sample are restricted to the period range of P$_0$ = 0.44\dots0.66~d which is only a subset of the accessible period range for these stars, as shown in figure~\ref{fig-4}. At longer periods only a single K2 star and a few bulge stars display modes that fall approximately into the $f_1$ group. This cut-off could be plausibly explained by the occurrence of the Blazhko effect itself. A recent study found that most modulated stars have pulsation periods below 0.7~d, and their mean period is 0.54 d \cite{2016A&A...592A.144S}. This is in perfect agreement with the distribution of the K2 stars with additional modes, even though the distribution of all RRab stars peaks at about 0.6 d.

However, modulated stars, although in lower numbers, extend to periods even shorter than 0.4~d. One prime example is V1127 Aql itself, the single occupant of the short-period end of figures~\ref{fig-1}--\ref{fig-2}. This lack of stars can be attributed the small numbers of short-period RRab stars observed by K2 in general so far: a larger sample could uncover similar stars from the K2 sample as well. 

The long-period OGLE stars were found in a study that specifically targeted bulge RRab stars with P$_0$ > 0.6 d. Although this criterion strongly limits the bulge sample, one would still expect an increase in the number of stars towards the cut-off. Yet, this not the case, hinting that the distribution of additional modes might be different for field and bulge RR Lyrae stars. 

\section{Conclusions}
New observations, especially from the K2 RR Lyrae Survey, have uncovered a plethora of low-amplitude additional modes in fundamental-mode RR Lyrae stars. These can be grouped into three broad categories established in recent years (namely, $f_1$, $f_2$, and PD), but our efforts to detect further structures in the Petersen diagram of these modes have not been successful so far. We suspect that multiple radial and non-radial modes are behind these groups and that temporal variation of the modes (either intrinsic or connected to the Blazhko effect) further complicate the identification of the modes. A more detailed study to incorporate more K2 campaigns and examine the amplitudes of the frequency peaks is currently under way. 

\section*{Acknowledgements}
Z.D., B.K., and J.O. wishes to acknowledge the hospitality of the Konkoly Observatory during their stay. This project has been supported by the NKFIH K-115709, PD-116175, and PD-121203 grants of the Hungarian National Research, Development and Innovation Office, and by the Lend\"ulet LP2014-17 Program of the Hungarian Academy of Sciences. The research leading to these results has received funding from the ESA PECS Contract No. 4000110889/14/NL/NDe, and from the European Community's Seventh Framework Programme (FP7/2007-2013) under grant agreement no. 312844 (SPACEINN). L.M. was supported by the J\'anos Bolyai Research Scholarship of the Hungarian Academy of Sciences.

%-----------------------------------------------------------------------
% BibTeX or Biber users please use (the style is already called in
% the class, ensure that the "woc.bst" style is in your local directory)
 \bibliography{LaszloMolnar}
%\begin{thebibliography}{}
%Journal Author, Journal \textbf{Volume}, page numbers (year)
%Book Author, \textit{Book title} (Publisher, place, year) page numbers
% etc
%\end{thebibliography}

%***********************************************************************
\end{document}